\documentclass[10pt]{article}
\usepackage[utf8]{inputenc}
\usepackage[T1]{fontenc}

\usepackage{graphicx}
\usepackage{pdfpages}

\usepackage{lipsum}
\usepackage{xcolor}

\usepackage{authblk}
\usepackage[margin=2cm]{geometry}
\usepackage{mathtools, cuted}

\usepackage{bm}
\usepackage{amsfonts}
\usepackage{amsmath}
\usepackage{amssymb}
\usepackage{mathrsfs} % cursive letters
\usepackage{siunitx}

\usepackage{soul}
\usepackage[normalem]{ulem}
\title{
Flat-cupped transition in freezing drop impacts}
\author[1]{Marion Berry}
\author[2]{Christophe Josserand}
\author[1]{Anniina Salonen}
\author[1]{François Boulogne}
\affil[1]{Laboratoire de Physique des Solides, UMR 8502, CNRS, Université Paris-Saclay, 91405 Orsay, France.}
\affil[2]{Laboratoire d’Hydrodynamique (LadHyX), UMR 7646 CNRS-Ecole Polytechnique, IP Paris, 91128 Palaiseau, France.}

\date{\today}
\begin{document}

%\twocolumn[
%    \begin{@twocolumnfalse}
        \maketitle
        \begin{abstract}
            We present an experimental study on the freezing of alkane drops impacted on a liquid bath.
            More specifically, for drops of hexadecane and tetradecane on brine, we found a morphological transition of the solid between a flat disk and a cupped shape.
            We show that this transition depends mainly on melting temperature and thermal shock, and varies weakly with impact velocity.
            We observed that the impact dynamics does not depend on the thermal shock before the drop starts to solidify, which allows a rationalization of the solid size by models established for impact without phase change.
            Finally, we show that the relevant timescale setting the onset of solidification is associated with the formation of a thin solid layer between the drop and the bath, a timescale much shorter than the total solidification time.
            These findings offer the possibility to collapse the data for both liquids in a single phase diagram.
        \end{abstract}
%    \end{@twocolumnfalse}
%]

%%%%%%%%%%%%%%%%%%%%%%%%%%%%%%
%
% INTRODUCTION
%
%%%%%%%%%%%%%%%%%%%%%%%%%%%%%%
\section{Introduction}

%\paragraph{Header\\}

Solidification processes are used for a wide range of applications, including metallurgy, spray cooling, 3D printing, or aircraft icing issues. The resulting frozen patterns are influenced by the detailed interaction between solidification and capillarity, leading to various shapes, from spherical ice drops to flat splats.
Indeed, the production of non-spherical objects has been studied through microfluidics and self-assembly processes, both requiring very specific equipment.
It is thus tempting to investigate how solidification could influence and control the formation of solid objects.
Among the different processes, solidification upon impact of liquid drops has recently received a growing interest, due to its simplicity and its importance on aeronautic applications under freezing conditions.

More generally, over the last years, impact dynamics has become an iconic problem of multiphase flows, primarily for its industrial and environmental interests, but also because it exhibits the main fundamental challenges of such flows. Impact can either occur on solid surfaces but also on liquid, sharing similar features as well as important differences, in particular when wetting is involved. It is thus interesting to investigate how these behaviors vary as the drop solidifies upon impact.

On solid surfaces, Thiévenaz \textit{et al.} \cite{Thievenaz2019,Thievenaz2020PRF} have observed that water droplets quickly spread and form an ice layer, on top of which the remaining liquid retracts.
They reported two morphologies, denoted cap shape and ring shape, resulting from the competition between the retraction of the liquid and the solidification time.
The variety of morphologies in the ring shape category has been further refined by Fang \textit{et al.} \cite{Fang2021}, who showed that in some cases two freezing fronts can advance simultaneously.
As the drop adheres to the cold substrate, the freezing process also affects the spreading of the solid layer, thus the maximum radius of the final morphology.
Gielen \textit{et al.} \cite{Gielen2020} and Thiévenaz \textit{et al.} \cite{Thievenaz2020EPL} studied different systems, molten tin drops and water drops respectively.
They both showed that the relation between the maximum spreading radius and usual impact parameters, well described for spreading upon isothermal impact, does not account for the solidification-limited spreading.
The experimental data are well explained by a model that considers an effective viscosity arising from the the combined  viscous and ice growth boundary layers.
In addition, Gorin \textit{et al.} explored the maximum spreading radius with different liquids and wettability of the solid substrates \cite{gorin2022}.
However, the morphologies accessible through impact on solid substrates remain limited.

Alternatively, impacts on liquids have been scarcely studied although they are a promising route for obtaining a large variety of solidified drop morphologies.
The richness of morphologies is due to the combination of physical parameters from both the drop and the bath.
In particular, the substrate thickness is relevant and usually classified as a liquid layer or a liquid bath through a dimensionless bath thickness $H$ defined as the ratio of the bath thickness to the drop diameter \cite{Macklin1976,Tropea1999,vander_wal_2006}.
In the present study, we work with a liquid bath, which mainly implies a cavity opening upon impact.

The dynamics of the cavity has been intensely studied in experiments, simulation and theories, in the absence of solidification.
In particular, Pumphrey and Elmore \cite{Pumphrey1990}, and Oguz and Prosperetty \cite{Oguz1990} both built a scaling to describe the maximum cavity radius, which is based on the balance between the drop kinetic energy and the potential energy of the displaced liquid associated to the opening of a cavity in the bath.
This scaling has been written for impact of miscible drops, specifically water on water impacts.
Liow \cite{Liow2001} refined this scaling by introducing a factor of energy conversion in the energy balance.
Through a fit on data from various water-on-water studies, he highlighted that 28~\% of the initial drop energy is converted to open the cavity.
This refined scaling is widely used as it appears to fit data from a wide range of systems, for instance with baths of varying viscosity using water-glycerol mixtures \cite{fedorchenko2004} and even with non miscible liquids, \textit{i.e.} viscous oil drops impacting a water bath \cite{Jain2019,Hasegawa2019}.
With similar assumptions, Berberovic \textit{et al.} \cite{berberovic2009} and Bisighini \textit{et al.} \cite{bisighini2010} model the time evolution of the cavity radius for water on water impacts.
No analytical solutions were found, however they compare their experimental data with numerical simulations, showing good agreement.
Finally, Che and Matar \cite{Che2018} systematically explored the role of the spreading factor through water on oil and oil on water impacts, as well as the role of substrate thickness and viscosity.
Their experimental work highlighted that immiscibility has a considerable effect on the impact process, illustrating its complexity and richness of behavior.

Therefore, solidification of drops impacting a liquid substrate offers an interesting playground to produce a wide range of morphologies.
Some works explored chemical reaction with alginate drops in calcium bath \cite{Lee2015,Chan2009}. % (bath concentration and cross linking rate)
Other works chose solidification process with molten wax drops \cite{Beesabathuni2015,Lee2015} or tin drops \cite{wang2021}.
Recently, Wang \textit{et al.} \cite{Wang2023} studied hexadecane drops with a cooled down water bath and compared solidification time and characteristic impact times to study the morphological transitions.

The aim of the present work is thus to characterize and identify the interplay between the interface dynamics and the solidification for alkane drop impact on a cold water bath. The paper is organized as follows.
In Section \ref{sec:methods-obs}, we present our experimental methodology, the main physical parameters of the problem, as well as experimental observations of the impact dynamics and final morphologies of the solid.
Next, in Section \ref{sec:dynamics}, we explore the impact dynamics and solidification process and discuss two different radii related to the problem.
Finally, in Section \ref{sec:shape}, we rationalize the transition between the two morphologies observed by comparing scaling laws for typical timescales.

%%%%%%%%%%%%%%%%%%%%%%%%%%%%%%
%
% EXPERIMENTAL METHOD
%
%%%%%%%%%%%%%%%%%%%%%%%%%%%%%%
\section{Methodology and observations}\label{sec:methods-obs}

%%%%%%%%%%%%%%%%%%%%%%%%%%%%%%
\subsection{Experimental methods}\label{ssec:exp-methods}

The experiment consists of releasing a drop of alkane at room temperature, $T_{\rm d}$, onto a liquid bath of NaCl brine at a temperature $T_{\rm b} \leq T_{\rm d}$.
The drop temperature $T_{\rm d}$ is measured for each experiment with a precision of $\pm 0.1~^\circ$C and varies between $19$ and $23~^\circ$C.
The two alkanes used in this study are tetradecane (purity > 99 \%, Sigma-Aldrich), of melting temperature $T_{\rm m,t}$ = 5.6~$^\circ$C, and hexadecane (purity > 99 \%, Sigma-Aldrich), $T_{\rm m,h}$ = 18.1~$^\circ$C.
The brine is made of NaCl (purity > 98 \%, Sigma-Aldrich) at 23.3 wt \% and pure water (Resistivity 18.2 MΩ·cm, Purelab Chorus, Veolia).
The bath is cooled down by a Peltier module (52 $\times$ 52mm, 35.8V, 15.4A, 340.5W, Radiospare) and a refrigerated circulator (model AP15R-40, VWR).
Its temperature $T_{\rm b}$ is measured with a K-type thermocouple (Radiospare) located near the surface of the bath and ranges from $T_{\rm d}$ to $-21$ $^\circ$C.
The drops are produced by a needle (18G, internal diameter 0.84 mm, Poly Dispensing) connected to a syringe pump (Pump 11 Elite, Harvard Apparatus) with a constant flow rate of 0.02 ml/min.
This produces drops of diameter $2r_0 = 2.50 \pm 0.04$~mm, measured by image analysis.
The dimensionless bath thickness $H$ is defined as the ratio of the bath thickness to the drop diameter.
In this study, the bath thickness is 10 mm, thus giving $H$ = 4, which is classified as liquid bath in the literature \cite{Macklin1976,Tropea1999,vander_wal_2006}.
The drop is released at an initial height $h_0$ ranging from 3 cm to 40 cm, and the impact velocity is calculated as $v_0 = \sqrt{2g h_0}$ where $g$ is standard gravity, which corresponds to a velocity range $v_0\in [0.8,2.8]$ m/s.
In our experimental conditions, the solidification time varies between one to one and a hundred of seconds, whereas the typical impact time is of the order of tens of milliseconds.

Visualizations are performed with two different cameras.
A Basler camera (160um pro) with a telecentric lens (TEC-M55 55mm, Computar), operating between 20 to 50 fps, records images from the top to have an overview of the solidification process.
The pictures from this camera are used to estimate the solidification time and measure the final projected radius $r_{\rm f}$.
This radius is calculated from the surface area assuming a circular shape.
A high-speed FASTCAM Nova camera (Photron) with a micro-nikkor lens (AF-S 105 mm, Nikon) records between 3000 to 9000 fps from a 3/4 point of view to have a more precise insight on the typical impact timescale, to measure the projected radius over time $r(t)$ and estimate the time $t_{\rm layer}$ to form a frozen layer.

%%%%%%%%%%%%%%%%%%%%%%%%%%%%%%
\subsection{Main physical parameters}\label{ssec:numbers}

Now, we introduce the parameters and dimensionless numbers used in impact dynamics and in solidification problems.
A subscript b refers to the bath properties, the liquid phase of the drop is associated with a subscript d, and the solid phase is denoted by the subscript s.

Regarding the solidification process, $C_p$ and $\lambda$ represent the specific heat capacity and thermal conductivity.
From these physical parameters, we define the thermal diffusivity,
\begin{equation}
    D_i = \frac{\lambda_i}{\rho_i C_{p,i}},
\end{equation}
which quantifies thermal transfers by conduction.
Another important parameter is the thermal effusivity, which measures the ability of a material to transfer thermal energy to its surroundings, written as
 \begin{equation}
     e_i = \sqrt{\lambda_i \rho_i C_{p,i}} = \frac{\lambda_i}{\sqrt{D_i}}.
 \end{equation}
Effusivity is the parameter to describe heat transfer when two materials at different temperatures are in contact.
The temperature at the interface $T_{\rm c}$ is constant in the case of two semi-infinite media and writes \cite{Boeker2011}:
\begin{equation}
    T_{\rm c} = \frac{T_{\rm d} + T_{\rm b} e_{\rm b}/e_{\rm d}}{1+e_{\rm b}/e_{\rm d}},
    \label{eq:Tcontact}
\end{equation}
for a liquid drop in contact with the bath.
The hypothesis of a semi-infinite phase for the drop is correct for timescales smaller than the heat diffusion timescale, which is typically  of the order of magnitude of the second for our typical length scale whereas the timescale of impact is tens of milliseconds.
To quantify the bath temperature $T_{\rm b}$, several temperature differences can be introduced.
The most used is the thermal shock defined as $\Delta T_{\rm mb} = T_{\rm m} - T_{\rm b}$, however another can be defined with the contact temperature, which will be called $\Delta T_{\rm mc} = T_{\rm m} - T_{\rm c}$.
A typical dimensionless number associated with phase change in a material is the Stefan number, defined with the enthalpy of solidification $\mathcal{L}$ and the specific heat capacity of the solid drop as
\begin{equation}
    {\rm St} = \frac{C_{p,s} \Delta T_{\rm mb}}{\mathcal{L}}.
\end{equation}

We additionally introduce three dimensionless numbers to characterize the impact dynamics.
The main parameters influencing drop impact are impact velocity $v_0$, initial drop radius $r_0$, the properties of the drop and the properties of the liquid bath.
The density ratio is defined as $\rho' = \rho_{\rm b} / \rho_{\rm d}$.
The Froude number quantifies the effect of gravity
\begin{equation}
    {\rm Fr} = \frac{v_0^2}{ 2 r_0\,g},
\end{equation}
and the Reynolds and Weber numbers compare the effect of inertia with, respectively, viscous and capillary forces,
\begin{equation}
    {\rm Re} = \frac{2 \rho_{\rm d} v_0 r_0}{\mu_{\rm d}},
\end{equation}
\begin{equation}
    {\rm We} = \frac{2 \rho_{\rm d} v_0^2 r_0}{\sigma_{\rm d}},
\end{equation}
where $\rho_{\rm d}$, $\mu_{\rm d}$, and $\sigma_{\rm d}$ are the density, dynamic viscosity, and surface tension of the liquid drop \cite{Jain2019}.

The values used for the thermal properties come from literature and the fluid properties are measured by ourselves as a function of their temperature.
For the calculations, the alkanes' properties in their liquid phase are taken at $T_{\rm d}$ = 20~$^{\circ}$C and in their solid phase at melting temperature $T_{\rm m}$.
As the bath temperature ranges from $T_{\rm d}$ to $-21$ $^\circ$C, the NaCl brine properties are taken at a mean temperature of around 0~$^{\circ}$C.
The results from our measurements are presented in SI, as well as a more detailed comment on the temperature dependency.

In our experiments, the ranges of the hydrodynamic numbers are  ${\rm Fr} \in [24,340]$, ${\rm Re} \in [570,2260]$, ${\rm We} \in [40,515]$.
These numbers mean that inertia dominates over viscous and capillary forces during the impact dynamics and that gravity does not play a role when the drop impacts the bath.
Also, the Stefan numbers are ${\rm St} \in [0.04,0.34]$ indicating a faster heat diffusion process than solidification dynamics.

%%%%%%%%%%%%%%%%%%%%%%%%%%%%%%
\subsection{Experimental observations}

\begin{figure*}[ht]
\centering
    \includegraphics[width=0.8\linewidth]{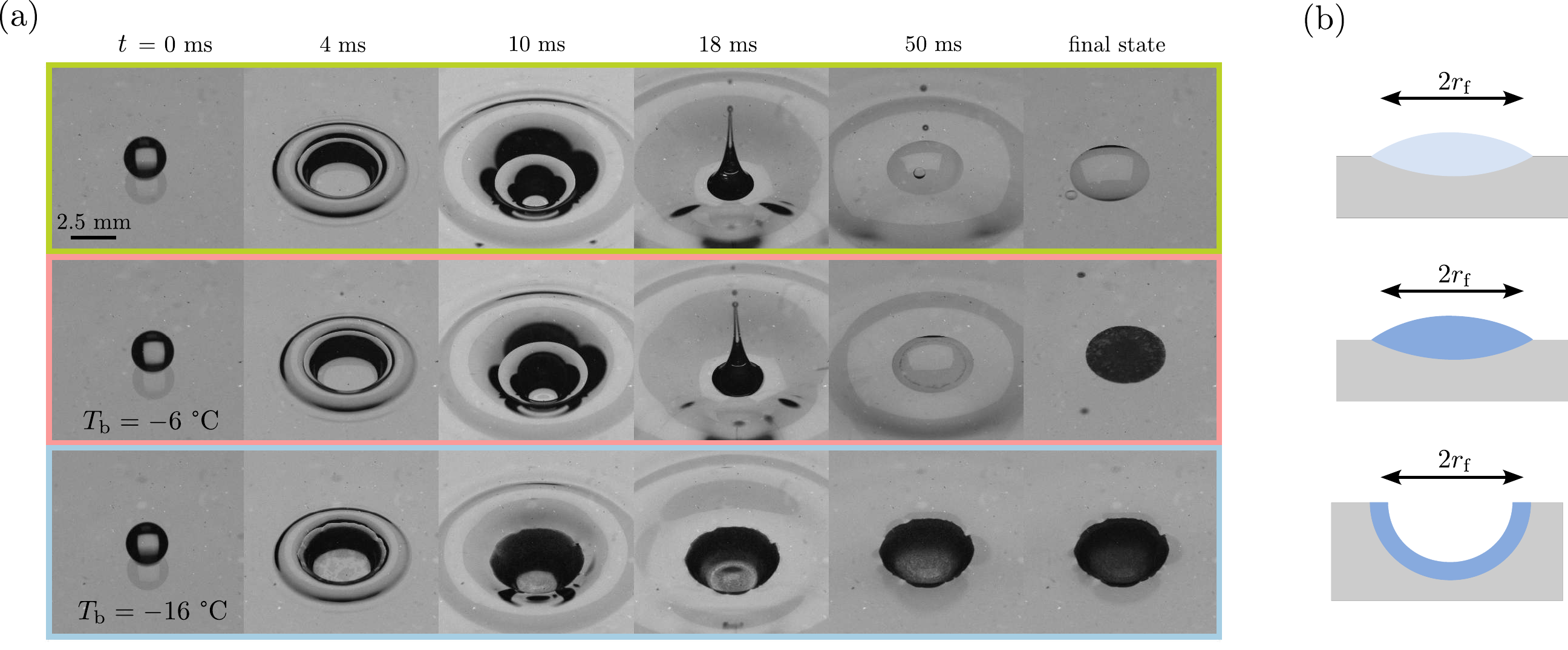}
\caption{(a) Time series showing a tetradecane drop impinging a liquid bath at a velocity $v_0$ = 1.4 m/s (Fr = 80, We = 134).  for three different bath temperatures $T_{\rm b} =\{T_{\rm d}, -6, -16\}~^\circ$C.
In the first row the drop remains liquid and forms a liquid lens at the surface of the bath.
The final state of the second and third rows corresponds to the fully solidified drop which occurs at about 50 s and 5 s, respectively.
Associated videos are provided in Supplementary Information.
(b) Side view schematics of the final shape. The thickness and angle at the contact line are not meant to be realistic. On the first row, the drop forms a liquid lens.
On the second and third rows, the drop is solidified in a disk-like shape and a cupped shape, respectively.
}
\label{fig:raw_data1}
\end{figure*}

Figure~\ref{fig:raw_data1}(a) illustrates typical experiments at a fixed impact velocity $v_0$ for three different bath temperatures $T_{\rm b} =\{T_{\rm d}, -6, -16\}~^\circ$C.
In the first row, for the isothermal impact, the impact opens a cavity up to a maximum radius $r_{\rm max}$.
Then, this cavity retracts and produces a vertical jet.
Finally, the system reaches an equilibrium where a flat liquid lens floats at the surface of the bath due to the immiscibility of and the density contrast between the two liquids.

In the second and third rows in figure \ref{fig:raw_data1}(a), the bath temperature is lower than the melting temperature of the drop, $T_{\rm b} < T_{\rm m}$.
We observe that the impact dynamics is similar to the first row until 10 ms and the cavity formed by the drop has the same width for the three bath temperatures.
In the second row, for a moderate bath temperature, the dynamic after 10 ms is exactly the same as at $T_{\rm d}$ with the cavity retraction and the ejection of a jet. The drop solidifies after impact, forming a flat solid disk as illustrated in figure~{\ref{fig:raw_data1}}(b).
In the last row, for a lower bath temperature, we notice an irregular contact line and gray clusters at 4 ms, which we attribute to crystals growing. The solidification has started but it does not appear to significantly modify the impact dynamics. At 10 ms, the drop is darker than the other rows due to solidification, indicating the formation of a solid layer in the drop. At 18 ms, the drop shows significant differences with the other rows, which we attribute to the solid layer stopping the impact dynamics in the cavity state. We still observe some liquid moving on top of the solid layer but no ejection of a jet.
In this case, a solid cup is obtained as illustrated in figure~{\ref{fig:raw_data1}}(b).
At 50 ms, the remaining liquid phase of both drops is still solidifying.
The final state shows the final solidified drops, which corresponds to a solidification time of about 50 s and 5 s, respectively.

\begin{figure*}[ht]
\centering
\includegraphics[width=0.8\linewidth]{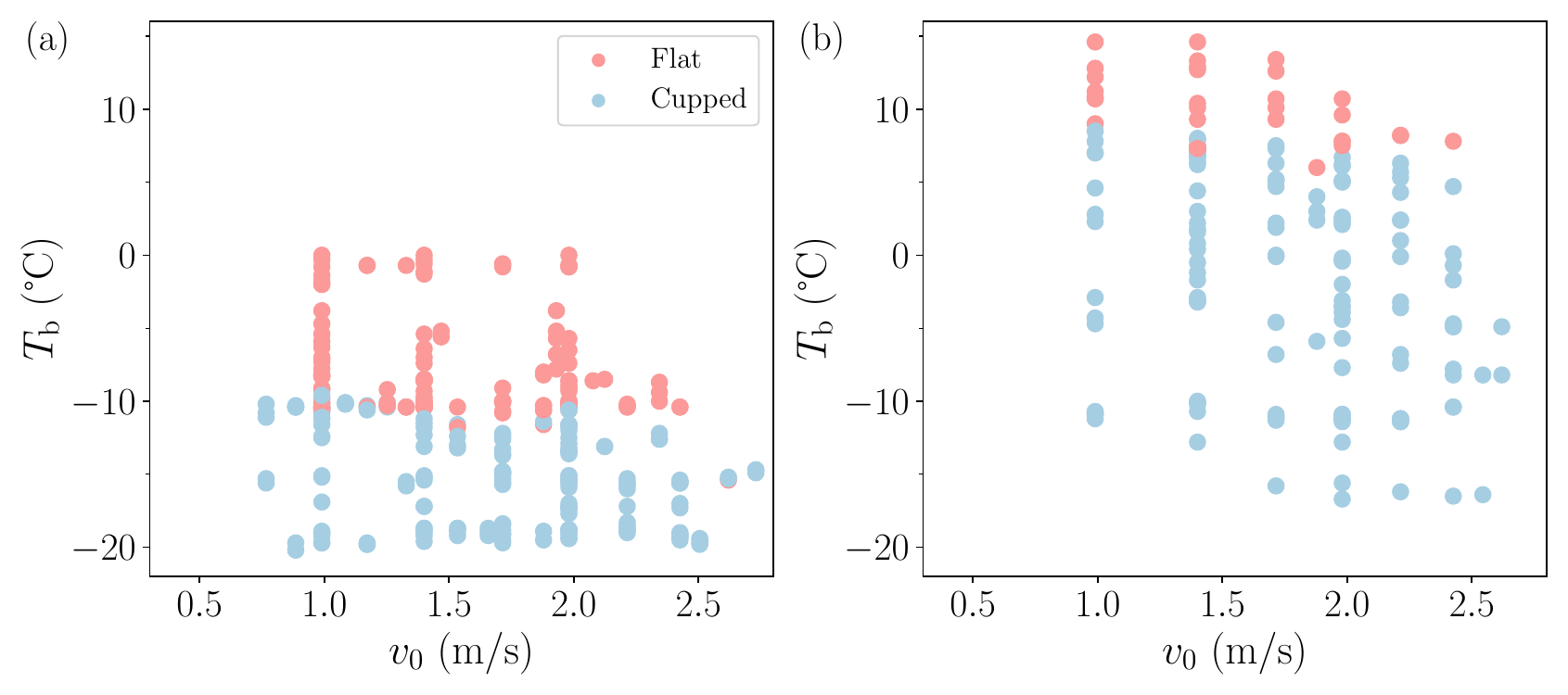}
\caption{(a) Phase diagram for drops of tetradecane representing the bath temperature $T_{\rm b}$ as a function of the impact velocity  $v_0$ where the two types of morphology are reported: flat or cupped.
(b) Phase diagram for drops of hexadecane.
}
\label{fig:raw_data2}
\end{figure*}

We have systematically varied the bath temperature $T_{\rm b}$ and the impact velocity $v_0$, and classified the final morphology of the solidified drop in two categories: flat or cupped.
The classification is made on visual inspection when the solidified drop is retrieved from the bath, with a criterion on concavity.
We report these results in figure~\ref{fig:raw_data2}(a) for tetradecane and  figure~\ref{fig:raw_data2}(b) for hexadecane.
Both liquids present a flat-cupped transition that is nearly independent of the drop velocity $v_0$.
As expected the bath temperature cannot be used to normalize the phase diagrams as the transition is at $T_{\rm b} \simeq -10$~$^\circ$C for tetradecane and $T_{\rm b} \simeq 6$~$^\circ$C for hexadecane.
Also, the usual thermal shock $\Delta T_{\rm mb}$, defined with the bath temperature, cannot capture the transition as it would predict different values for the two systems, namely 16~$^\circ$C for tetradecane and $10$~$^\circ$C for hexadecane.
However, a thermal shock defined with the contact temperature gives a transition of about $\Delta T_{\rm mc} \simeq 8$~$^\circ$C for both liquids.
Therefore, $\Delta T_{\rm mc}$ will be used in the following to combine the results of both liquids.

%%%%%%%%%%%%%%%%%%%%%%%%%%%%%%
%
% RESULTS & DISCUSSION
%
%%%%%%%%%%%%%%%%%%%%%%%%%%%%%%
\section{Impact dynamics and solidification process}\label{sec:dynamics}

%%%%%%%%%%%%%%%%%%%%%%%%%%%%%%
\subsection{Dynamics of the cavity}

\begin{figure}[ht]
\centering
    \includegraphics[width=.8\linewidth]{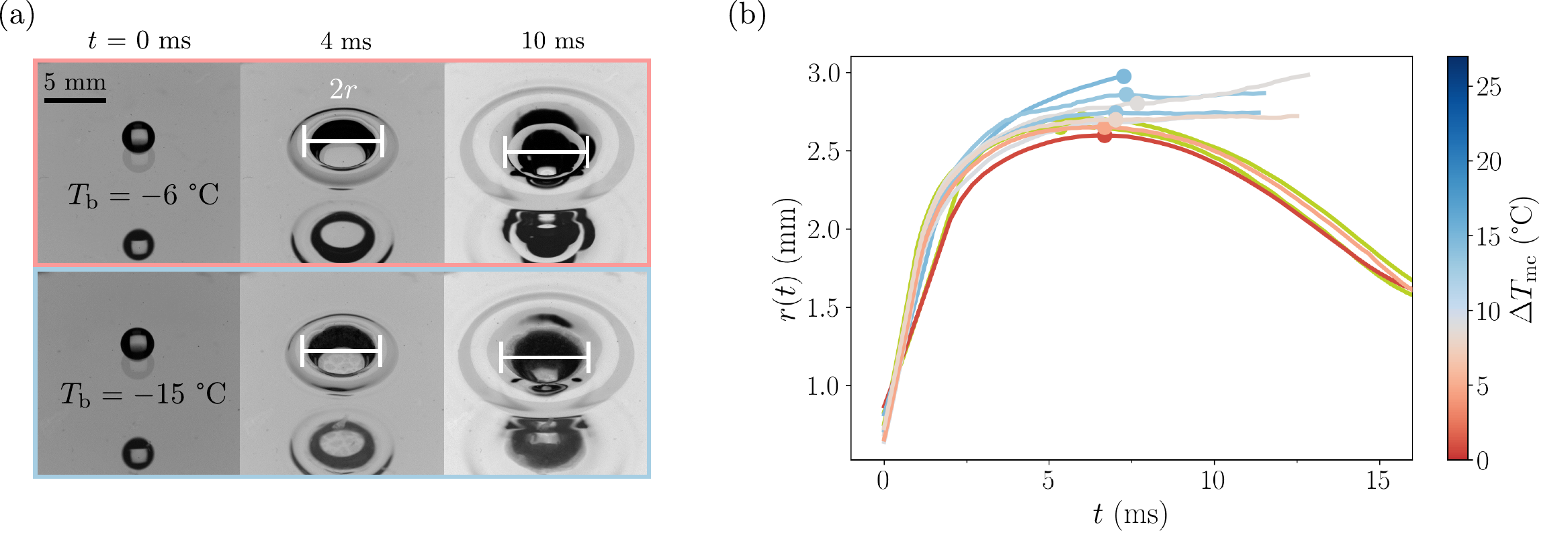}
\caption{(a) Time series illustrating the measurement of the drop radius $r(t)$.
(b) Dynamics of the cavity radius $r(t)$ for drops of tetradecane for $v_0$ = 1.4 m/s (Fr = 80).
The green lines represent isothermal impact.
The color of the other lines corresponds to the thermal shock $\Delta T_{\rm mc}$ indicated by the color bar.
The points represent $(t_{\rm max}$, $r_{\rm max})$ associated with each experiment.  }
\label{fig:dynamics-cavity-radius}
\end{figure}

Now, we focus on the description of the opening dynamics of the cavity.
For the sake of simplicity, we report the cavity radius as measured on the drop.
In figure~\ref{fig:dynamics-cavity-radius}(a), we show  the opening of the cavity for two different thermal shocks $\Delta T_{\rm mc}$.
Measurements of the cavity radius for drops of tetradecane at a constant impact velocity are reported in figure~\ref{fig:dynamics-cavity-radius}(b).
The green lines represent two isothermal impacts, $T_{\rm b} = T_{\rm d}$, showing a good reproducibility of our measurements and data.
We observe an increase of the cavity width reaching a maximum radius $r_{\rm max}$, which is eventually followed by a retraction of the cavity.

Even when the bath is cooled down, the beginning of the impact dynamics is similar to the case of an isothermal impact.
We identify two behaviors: small $\Delta T_{\rm mc}$ for $ [0,8]$~$^\circ$C and larger $\Delta T_{\rm mc}$ above 8~$^\circ$C.
When $\Delta T_{\rm mc}$ is rather small (red lines), the radius evolution is the same as the isothermal experiments.
We observe the closing of the cavity, which leads to a liquid lens  and a flat shape as a final state.
When $\Delta T_{\rm mc}$ is large enough (blue lines), the cavity radius stays constant or slightly increases after reaching $r_{\rm max}$, which leads to a cupped shape as a final state.

The temporal evolution of the radius is asymmetric, and opening takes about 6 ms while retraction takes more than 15 ms in figure~\ref{fig:dynamics-cavity-radius}.
This delay in the retraction is attributed to a loss of energy by viscous dissipation during impact.
Most models do not include the viscous dissipation and thus describe a symmetrical opening and closing of the cavity.
Nevertheless, predictions for the maximum radius and the time associated are available in the literature for impact of a drop in a bath of the same liquid \cite{Pumphrey1990,Oguz1990,Liow2001}.
We thus consider these predictions to describe our measurements in the next paragraph.

%%%%%%%%%%%%%%%%%%%%%%%%%%%%%%
\subsection{Maximum cavity radius}

\begin{figure}[tb]
\centering
    \includegraphics[width=.49\linewidth]{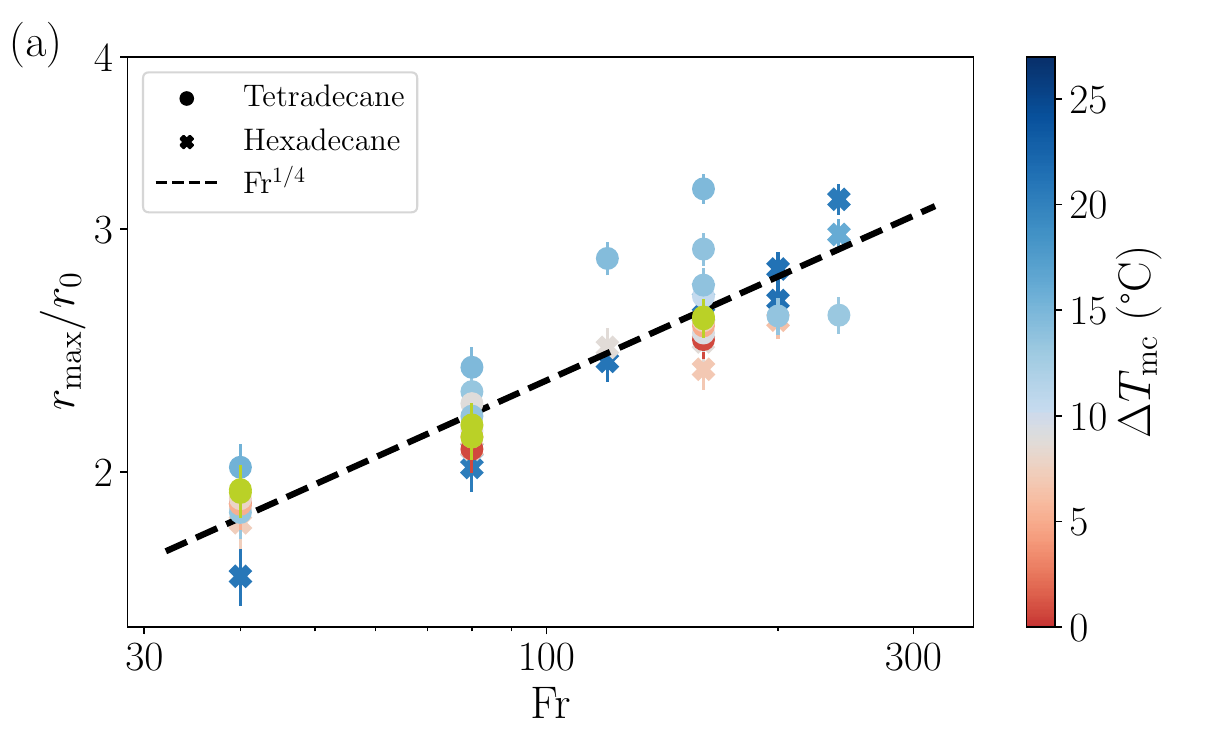}
    \includegraphics[width=.49\linewidth]{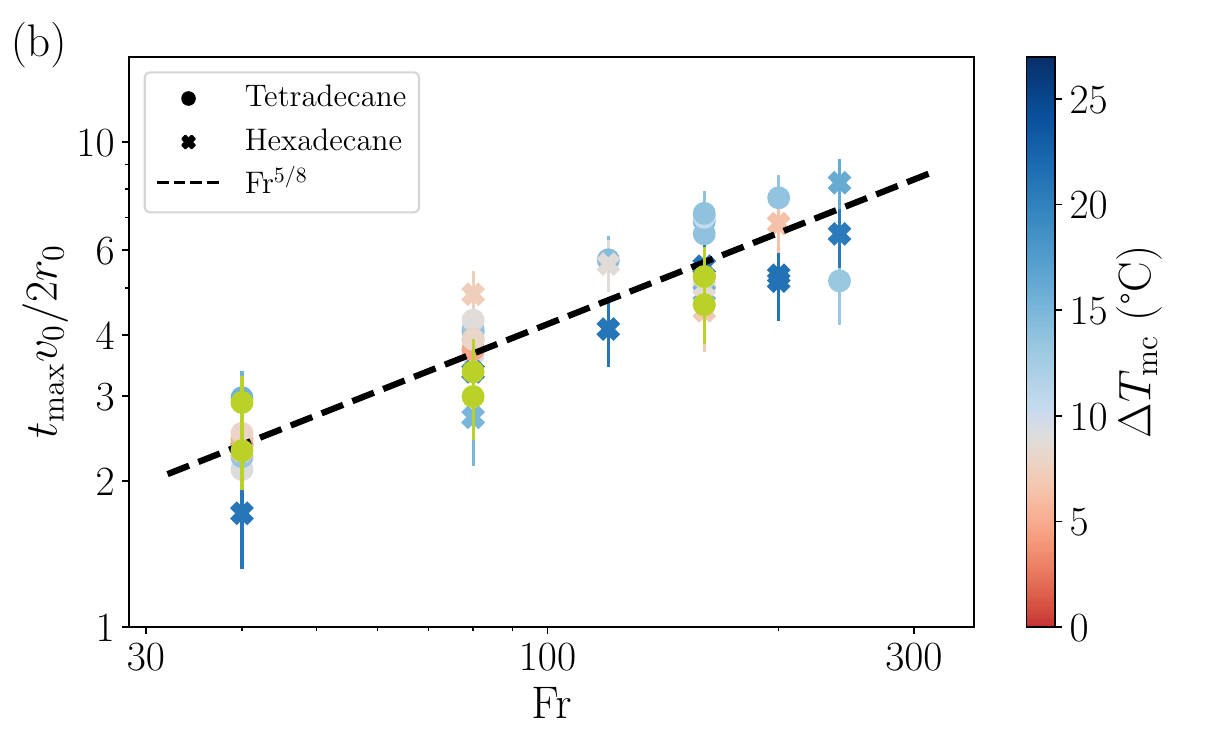}
\caption{(a) Dimensionless maximum radius $r_{\rm max}$$/r_0$ as a function of the Froude number in logarithmic scale. Each point represents one experiment. The green points are the cavity radius for isothermal impact. The dashed black line is equation \ref{eq:Rmax} with $\chi= 0.08$.
(b) Dimensionless timescale to reach the maximum radius $t_{\rm max}$$v_0 / 2r_0$  as a function of the Froude number in logarithmic scale.
On both plots, each point represents one experiment and the errorbars are determined using the variability of the corresponding $r_{\rm max}$.
The green points represent the isothermal impact and the dashed black line is equation \ref{eq:tmax} with $\xi= 0.23$.
}
\label{fig:Rmax-tmax}
\end{figure}

Pumphrey and Elmore, as well as Oguz and Prosperetti proposed to describe the maximum cavity depth with an energy balance between the drop kinetic energy and the potential energy associated with the displaced liquid in the bath \cite{Pumphrey1990,Oguz1990}.
To perform the calculation, the shape of the cavity is assumed to be hemispherical.
To obtain a better quantitative agreement with experimental results, Liow introduced later a factor $\chi$ ranging between 0 and 1 to consider that only a fraction of the kinetic energy is used in the cavity opening \cite{Liow2001}.
Thus, the energy balance leads to
\begin{equation}
    \frac{r_{\rm max}}{r_0} = 2 \left( \chi \frac{{\rm Fr}}{3\rho'} \right)^{1/4},
    \label{eq:Rmax}
\end{equation}
which has also been verified for a large variety of systems, including non-miscible liquids \cite{Jain2019,Hasegawa2019}.
Based on similar arguments, the timescale $t_{\rm max}$  to reach the maximum radius can be written as

\begin{equation}
    t_{\rm max} \frac{v_0}{2 r_0} = \xi \, {\rm Fr}^{5/8},
    \label{eq:tmax}
\end{equation}
where $\xi$ is a dimensionless factor \cite{Liow2001}.

In figure \ref{fig:Rmax-tmax}, we compare these predictions to our experimental results for both liquids with different impact velocities and thermal shocks.
Typically, the maximum radius $r_{\rm max}$ is in range $[2,4]$ mm and the timescale $t_{\rm max}$ in $[4,10]$ ms.
In figure \ref{fig:Rmax-tmax}(a), the experimental data for $r_{\rm max}$ normalized by the initial radius $r_0$, for different thermal shocks, agree with the scaling.
The fit of our data allows to obtain a good estimate of the cavity radius from the scaling law of Fr$^{1/4}$, which is widely used in the literature.
The value of $\chi$, 0.08, means that in this system 8~\% of the drop kinetic energy is used to form the cavity.
In Liow \cite{Liow2001}, the fitted value of $\chi$ is 28~\%, which is also used in other drop impact studies \cite{Jain2019,Hasegawa2019,engel1967}.
The model describes the opening of a cavity in the bath, however in the present study the cavity radius is measured on the drop.
As the drop does not cover the entire surface of the cavity (figure \ref{fig:dynamics-cavity-radius}(a)) due to surface energy cost, we underestimate the value of the cavity radius predicted by the model and the original energy conversion, thus explaining the difference of $\chi$.
A few experiments from the side were performed and the measured cavity depth data fitted with equation \ref{eq:Rmax} gives a factor $\chi$ of about 24~\%, which is closer to the usual value.
In figure \ref{fig:Rmax-tmax}(b), the experimental data for $t_{\rm max}$ normalized by the characteristic time $v_0/r_0$ are plotted for different thermal shocks.
The uncertainty on $t_{\rm max}$ is due to the shape of the $r(t)$ curve, even a small uncertainty on the measure of $r_{\rm max}$ causes a large uncertainty on $t_{\rm max}$.
Our data agree well with the Froude number to the power 5/8 suggested by equation~{\ref{eq:tmax}}, with a fitting parameter of $\xi$ = 0.23.

Thus, equations \ref{eq:Rmax} and \ref{eq:tmax} provide good estimates to describe the maximum opening of the cavity.
It therefore suggests that the expansion of the cavity is dominated by the inertia-gravity balance and that the solidification has only little influence on it.
In the next paragraph, we study the time necessary to reach the final state of the system, \textit{i.e.} the full solidification of the drop.

% %%%%%%%%%%%%%%%%%%%%%%%%%%%%%%
\subsection{Final radius and solidification time}\label{ssec:solidification_time}

\begin{figure}[h]
    \centering
    \includegraphics[width=\linewidth]{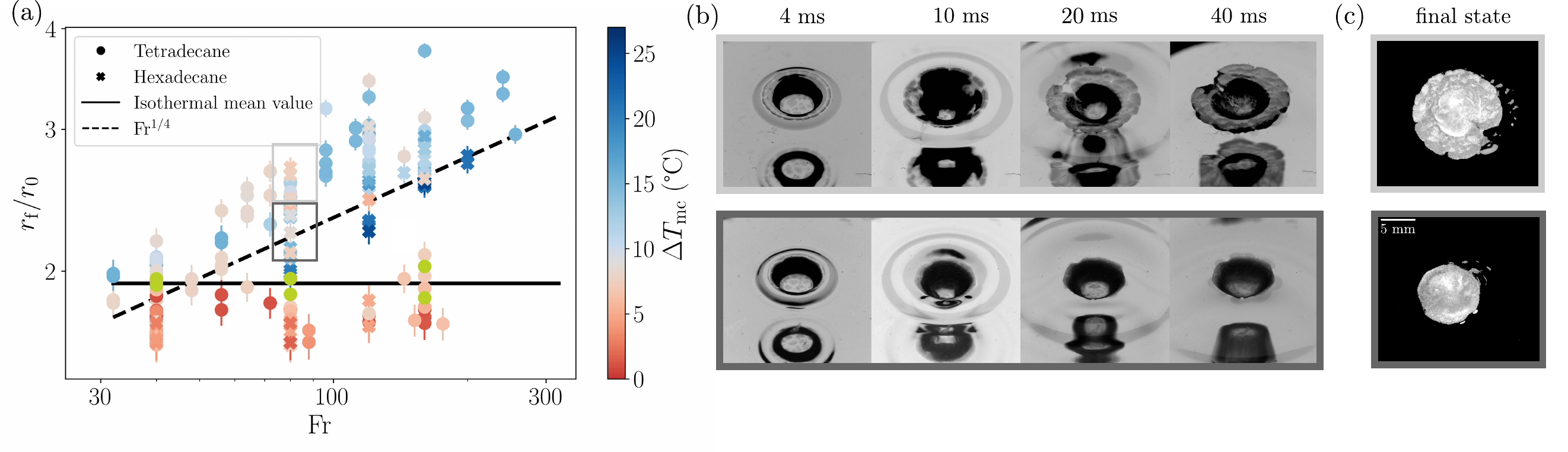}
    \caption{(a) Projected final radius $r_{\rm f}$ of the fully solidified drop as a function of the Froude number.
    Each point represents an experiment.
    The blue points correspond to solidification during impact.
    The dashed line represents $r_{\rm max}$ through equation \ref{eq:Rmax} with $\chi= 0.08$.
    The images show the different final morphologies and illustrate the deviation from the dashed line.
    The red points correspond to solidification after impact.
    The horizontal solid line represents $r_{\rm f} = 2.4$~mm, which corresponds to the average value at room temperature (green points).
    (b) Image sequence of tetradecane drops illustrating the scatter in the blue points for the same impact velocity. In the top image sequence the cupped shape is flattened after reaching $r_{\rm max}$.
    (c) Pictures from the top camera used for the measurements of $r_{\rm f}$. The presence of fractures is taken into account by taking the square root of the area, however the flattened part of the drop plays a significant role in the measured value, leading to an apparent final radius $r_{\rm f}$ larger than the maximum cavity radius $r_{\rm max}$.
    }
    \label{fig:Rfinal}
\end{figure}

The apparent final radius of the solidified drop is measured on the pictures from the top camera and plotted in figure \ref{fig:Rfinal}.
This radius depends drastically on thermal shock, which results in two distinct dependencies.

%%% Fig 4a

The blue points in figure \ref{fig:Rfinal}(a) correspond to large thermal shock where the drop solidifies with a cupped morphology.
As the final radius follows the scaling $r_0\,{\rm Fr}^{1/4}$, we can reasonably say that in this regime the drop solidified as a cavity of projected radius $r_{\rm max}$.
Nevertheless, we notice that most data points lie slightly above the prediction from  equation \ref{eq:Rmax} with the prefactor $\chi = 0.08$ determined in figure \ref{fig:dynamics-cavity-radius}.
Indeed, after reaching $r_{\rm max}$ and once a solid layer has grown on the drop, the system still has the energy associated to the retraction of the cavity.
This eventually leads to fractures spreading on the solid layer or a flattening of the solid layer, as illustrated in figure {\ref{fig:Rfinal}}(b). The final radius $r_{\rm f}$ is extracted from top camera pictures (figure {\ref{fig:Rfinal}}(c)) from the square root of the measured area.  The presence of fractures is absorbed in this estimation, however the flattened part of the drop plays a huge role in the measure.
Consequently, the apparent final radius $r_{\rm f}$ can be larger than the maximum cavity radius $r_{\rm max}$).

In addition, the transition between the two regimes (gray points) is at $\Delta T_{\rm mc} \simeq $ 8~$^\circ$C, which is consistent with the observed transition in the previous section.
Therefore, the visual change in morphology, flat or cupped, is mirrored in the evolution of $r_{\rm f}$.

%%% Fig 4b

In figure \ref{fig:Rfinal}(a), the green points are from an isothermal impact leading to a liquid lens of alkane at equilibrium at the surface of the bath, the size of which only depends on surface tension.
The gap between the points at Fr = 160 is too large to be associated with the uncertainty on the initial volume of the drop.
At this point, the Froude number is large enough to create a jet, which destabilizes into droplets \cite{Michon2017,Dhuper2021}.
The droplets ejected from the jet do not systematically fall back onto the initial drop.
It modifies the volume of the liquid lens at equilibrium, thus influencing the final radius.

The red points correspond to a small thermal shock and the final radius is constant regardless of the impact velocity.
This agrees with the observation made in figure \ref{fig:raw_data1} that if the drop solidifies after impact the radius $r_{\rm f}$ is close to the radius of a liquid drop at equilibrium.
The final radius is slightly below the value for isothermal impacts.
We attribute this difference to the temperature dependence of the liquid properties, which changes the equilibrium radius of the liquid lens.

% %%%%%%%%%%%%%%%%%%%%%%%%%%%%%%
\subsection{Discussion}

Two different radii are relevant for the freezing of drops impacting a liquid bath: the maximum cavity radius $r_{\rm max}$ and the radius of the final morphology $r_{\rm f}$.
For the bath temperatures explored in this work, the maximum cavity radius $r_{\rm max}$ reached the value observed in isothermal drop impact, which allowed us to use the proportionality $r_{\rm max} \sim r_0 \,{\rm Fr}^{1/4}$.
This observation is different from what happens during the freezing of drops impacting or deposited onto a solid substrate, in which freezing inhibits the observation of a maximum radius \cite{Gielen2020,Grivet2022}.
The final morphology of impacted drops on solid substrates is strongly coupled with the solidification time of the drop \cite{Thievenaz2020PRF,Fang2021,Meng2022}.
On liquid baths, the final morphology of the drop is not correlated with the solidification time, which was not highlighted in previous works.

%%%%%%%%%%%%%%%%%%%%%%%%%%%%%%
%
% RESULTS & DISCUSSION
%
%%%%%%%%%%%%%%%%%%%%%%%%%%%%%%
\section{Transition driven by interfacial crystallization}\label{sec:shape}
As shown in Section~\ref{sec:methods-obs}, the flat-cupped transition is mainly characterized by the thermal shock.
From the order of magnitude of the maximum opening time $t_{\rm max}$ ranging from 4 to 10 ms and the solidification time reported in section~\ref{ssec:exp-methods} larger than 1 s, the transition cannot be explained by a comparison of these two timescales.
Thus, we must consider the solidification process at shorter timescales.

In this Section, we model the formation of a solid layer between the drop and the bath, which is of the order of the millisecond.
We show that the flat-cupped transition is rather explained by comparison between the timescale associated to the formation of a thin solid layer, and that of impact dynamics.

%%%%%%%%%%%%%%%%%%%%%%%%%%%%%%
\subsection{Timescale for the formation of a frozen layer}

\begin{figure}
\centering
    \includegraphics[width=\linewidth]{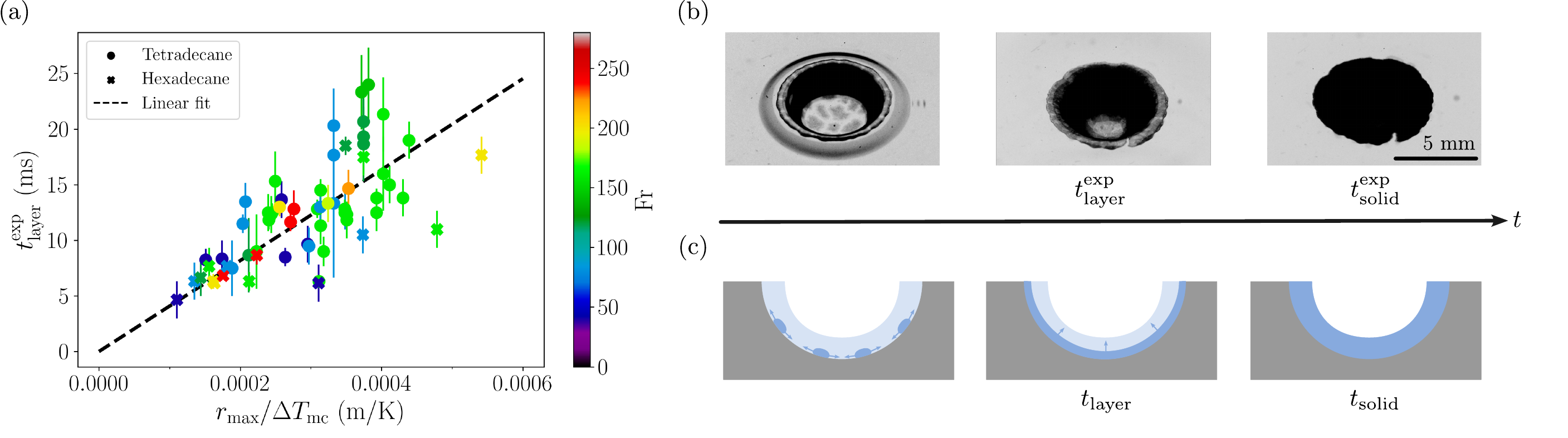}
\caption{ (a) Experimental estimation of the time $t_{\rm layer}^{\rm exp}$ to form a first layer as a function of $r_{\rm max} /  \Delta T_{\rm mc}$ for the two liquids and varying Froude numbers.
Each point corresponds to an experiment and the errorbar reflects the uncertainty associated with the observation.
The dashed black line is a linear regression of coefficient $1/(\sqrt{N_c}G) = 41$ K$\cdot$s$\cdot$m$^{-1}$  from equation~\ref{eq:tlayer}.
(b) Pictures illustrating how $t_{\rm layer}^{\rm exp}$ is estimated, for a bath temperature of $\Delta T_{\rm mc} \simeq 13$~$^\circ$C and an impact velocity of $v_0 \simeq  2$ m/s.
(c) Schematic representation of the solidification steps.
The crystals are growing at the interface between the drop and the bath, leading to the solid layer indicated by $t_{\rm layer}$, which then propagates in the bulk until full solidification of the drop is reached.
The schematic is not to scale for the sake of legibility.
}
\label{fig:first_layer}
\end{figure}

Once the drop is in contact with a bath at a lower temperature, the temperature of the drop in the vicinity of the interface between the two liquids decreases to reach the contact temperature $T_{\rm c}$.
For $\Delta T_{\rm mc} >0$, crystals nucleate and grow to form a solid layer along the drop/bath interface.
Due to the  temperature gradient that exists between the interface and the drop, the growth of crystals in a direction perpendicular to the interface is limited at this stage.
Then, the freezing front propagates towards the drop-air interface as described in \cite{Thievenaz2019}.

Let us consider a number $N_c$ of crystals growing in the drop at the interface with the bath.
The time to form a first layer $ t_{\rm layer}$ is reached when the crystals cover the entire surface area of contact.
The temperature at the interface is defined by equation \ref{eq:Tcontact} and when $\Delta T_{\rm mc}$ is positive, we consider that the drop at the interface is undercooled.
Thus, the crystals are growing at a constant velocity,
\begin{equation}
    v_{\rm cryst} = G \Delta T_{\rm mc},
\end{equation}
with $G$ the kinetic undercooling constant \cite{Pollatschek1929,Volmer1931,Hillig1956}.
De Ruiter \textit{et al.} and Koldeweij \textit{et al.} studied crystal growth in hexadecane and they measured experimental values for $G$ of $1.1 \times 10^{-2}$ and $ 0.45 \times 10^{-2}$ m/s/K, respectively \cite{de_ruiter2017,Koldeweij2021}.
Considering that the surface area to cover by the crystals scales as $r_{\rm max}^2$, the time scale to form the first solid layer is

\begin{equation}
    t_{\rm layer} = \frac{1}{\sqrt{N_c}}\frac{r_{\rm max}}{ G  \Delta T_{\rm mc}}.
    \label{eq:tlayer}
\end{equation}

To verify this prediction, we measured the time associated with the formation of the first layer $t_{\rm layer}^{\rm exp}$ in figure {\ref{fig:first_layer}}(a).
The formation of the solid layer can be observed on high speed camera movies as shown in figure {\ref{fig:first_layer}}(b) and illustrated in figure {\ref{fig:first_layer}}(c).

Then, we fit the data presented in  figure \ref{fig:first_layer}(a) with equation \ref{eq:tlayer}, in which $1/(\sqrt{N_c}G)$ is a fitting parameter.
From our data, we obtain $1/(\sqrt{N_c}G) = 41$ K$\cdot$s$\cdot$m$^{-1}$ and we do not observe any correlation with the impact velocity.
Since we observe about 10 to 20 crystals in our experiments, we can estimate an order of magnitude of the kinetic constant  $ G = [0.6 \pm 0.2] \times 10^{-2}$ m/s/K.
This value of $G$ is similar to the values measured in previous works on hexadecane \cite{de_ruiter2017,Koldeweij2021}, so this model is consistent with our data.

%%%%%%%%%%%%%%%%%%%%%%%%%%%%%%
\subsection{Flat-cupped transition}

\begin{figure}
    \centering
    \includegraphics[width=.55\linewidth]{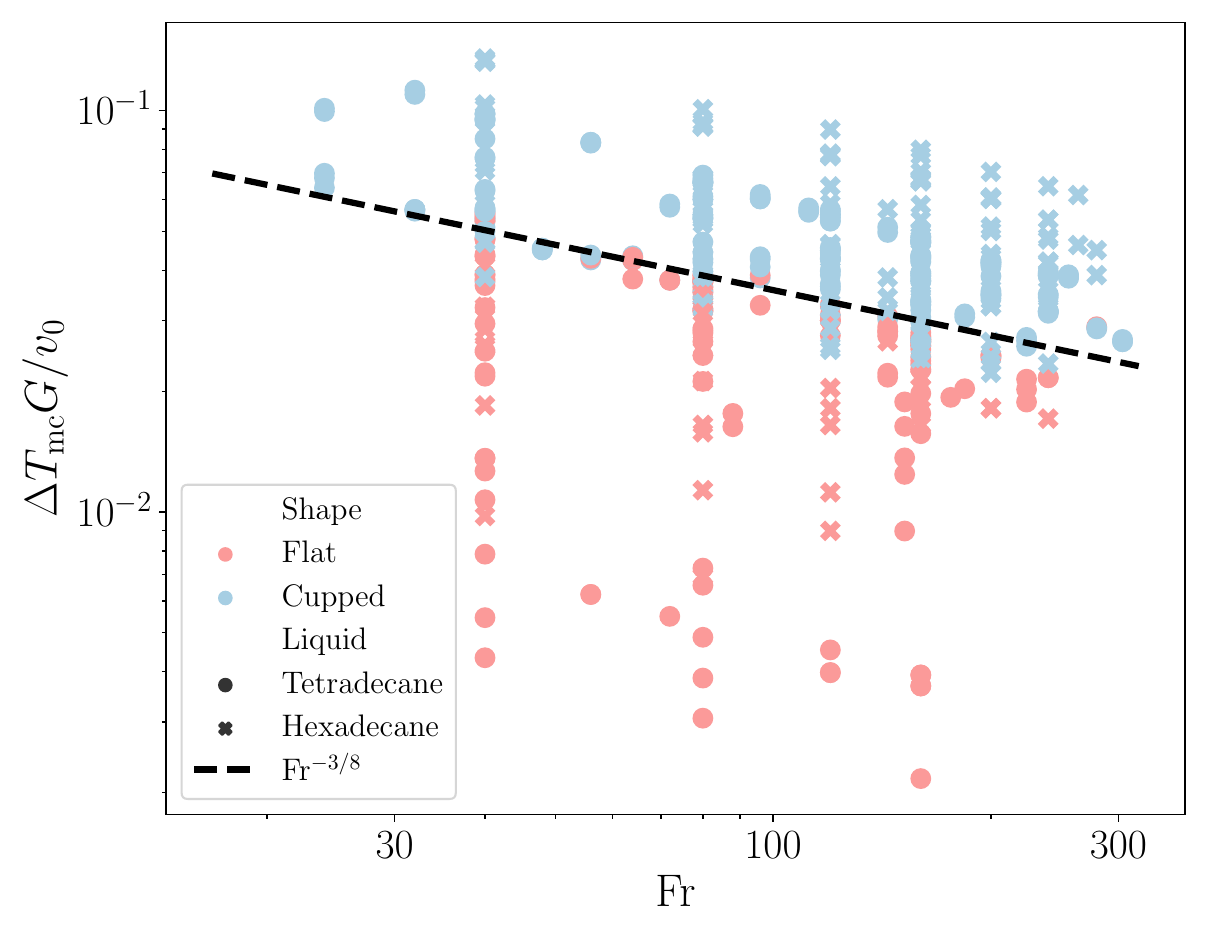}
    \caption{Phase diagram showing the dimensionless thermal shock $\Delta T_{\rm mc} G /v_0$ as a function of the Froude number in logarithmic scale, where the morphology of both liquids is represented.
    The dashed line corresponds to equation \ref{eq:transition} with a prefactor equal to 1/2.
    }
    \label{fig:normalized_phase_diagram}
\end{figure}

We anticipate that the formation of a solid shell beneath the drop on a timescale shorter or comparable to the impact duration can lead to a final concave solid.
Thus, we propose a criterion for the flat-cupped transition that writes  $t_{\rm max} \sim t_{\rm layer}$, which gives
\begin{equation}\label{eq:transition}
        \frac{\Delta T_{\rm mc} G}{v_0}  \sim {\rm Fr}^{-3/8}.
\end{equation}
This prediction suggests that the thermal shock of the transition weakly depends on impact velocity as $v_0^{1/4}$, which is compatible with the measurements presented in figure~\ref{fig:dynamics-cavity-radius}(c,d).
For a quantitative comparison, we plot a phase diagram for both liquids where the dimensionless thermal shock is plotted against the Froude number in figure~\ref{fig:normalized_phase_diagram}.
The predicted transition (\ref{eq:transition}) is plotted by a black dashed line with a fitting prefactor of 1/2, which describes well the transition.
The prefactor of the order of unity is coherent with our approach based on a scaling analysis and we note that the fitting parameter is the same for both liquids, which is a good indicator of the validity of the model.

%%%%%%%%%%%%%%%%%%%%%%%%%%%%%%
%
% CONCLUSION
%
%%%%%%%%%%%%%%%%%%%%%%%%%%%%%%

\section{Conclusion}

We considered the system of a drop of alkane impacting on a liquid bath, which is at a temperature lower than the melting temperature of the drop.
As a result, the drop solidifies and two final morphologies can be observed: flat or cupped.
This morphology depends in particular on the impact velocity and the bath temperature.

Experimentally, we observe that the maximum radius of the impact cavity and the associated duration of cavity opening are not affected by the solidification process.
In our conditions, the timescale of cavity opening is 10 ms, which is much shorter than the duration of freezing spanning between 1 to 100 s depending on the thermal shock and impact velocity.
This solidification time is found to be always much longer than the impact timescale, and is therefore not the relevant parameter to describe the flat-cupped transition.

Indeed, the morphological transition is attributed to a solidification process concomitant with the impact for the cupped morphology, and after impact for the flat one.
We have shown that the relevant thermal shock is defined as the difference between the drop temperature and the contact temperature between the drop and the bath.
The flat-cupped transition is explained by comparing the timescale associated with the dynamics of the cavity and the timescale for the appearance of a solid layer at the interface between the drop and the bath.
In particular, we demonstrate that the thermal shock associated to the transition varies as $v_0^{1/4}$.
This model perfectly describes the transition for two different alkanes of different melting temperatures for an impact velocity varying by an order of magnitude.
The model includes a phenomenological parameter $N_{\rm c}$, the number of growing crystals, that we considered as a constant for our entire dataset.
A theoretical description of this quantity is particularly challenging and the focus of recent studies \cite{kant2020,Koldeweij2021,Grivet2022}.
Our model would profit from future developments on crystal nucleation at interfaces.

Our findings open new perspectives for the production of non spherical objects, which are possible with the out of plane deformation induced by the liquid bath.
Extending this research to solidification processes relying on a mechanism different to freezing, such as chemical reaction, will be the purpose of future studies.

\section*{Acknowledgments}

This work received a financial support from  Investissements d’Avenir, LabEx PALM (ANR-10-LABX-0039-PALM).
We are grateful to Laura Wallon for performing the measurements of the fluid properties, Stéphane Cabaret for designing the heat exchanger, Alice Requier and Rodolphe Grivet for useful discussions.

\bibliography{biblio}

\bibliographystyle{unsrt}

% for arxiv
\newpage\clearpage


\section*{Supplementary material}

# Supplementary information: Flat-cupped transition in freezing drop impacts

Marion Berry[1], Christophe Josserand[2], Anniina Salonen[1], and François Boulogne[1]

[1]Université Paris-Saclay, CNRS, Laboratoire de Physique des Solides, 91405, Orsay, France.
[2]Laboratoire d'Hydrodynamique (LadHyX), UMR 7646 CNRS-Ecole Polytechnique, IP Paris, 91128 Palaiseau, France

## 1 Physical properties of materials

The physical properties of the bath, NaCl 23.3 % wt. brine, and the two alkanes, tetradecane and hexadecane, are detailed in this section. A subscript b refers to the bath properties, the liquid phase of the drop is associated with a subscript d, and the solid phase is denoted by the subscript s. Here, we combine values taken from the literature and measurements performed by ourselves.

### 1.1 Measured properties

This paragraph describes the methods and the results of our measurement of the fluid properties as a function of temperature.

- Liquid densities are measured with a densimeter from Anton Paar (DMA 5000 and 5001, temperature range: [0, 100] °C).

- Liquid viscosities are measured with the rheometer MCR 302, Anton Paar using a Couette geometry. The temperature is controlled through a JULABO F12-ED fluid circulator ([−20, 100] °C)

- The surface tension is measured on a Tracker from Teclis and computed with the previously measured densities. The liquid-vapor measurements are performed with the pendant drop method, taken over 60 seconds for each temperature. The surface tension stays constant during this duration, thus a mean value is taken. The liquid-liquid values are obtained with two methods, the pendant drop and the rising drop.

The variations of these three properties with the temperature are presented in figure S1.

### 1.2 Computed thermal properties

The thermal diffusivity quantifies thermal transfers by conduction inside a medium,

$$D_i = \frac{\lambda_i}{\rho_i C_{p,i}}, \tag{S1}$$

with $C_{p,i}$ and $\lambda_i$ the specific heat capacity and thermal conductivity respectively. The thermal effusivity measures the ability of a material to transfer thermal energy to its surroundings. It is written as

$$e_i = \sqrt{\lambda_i \rho_i C_{p,i}} = \frac{\lambda_i}{\sqrt{D_i}}. \tag{S2}$$

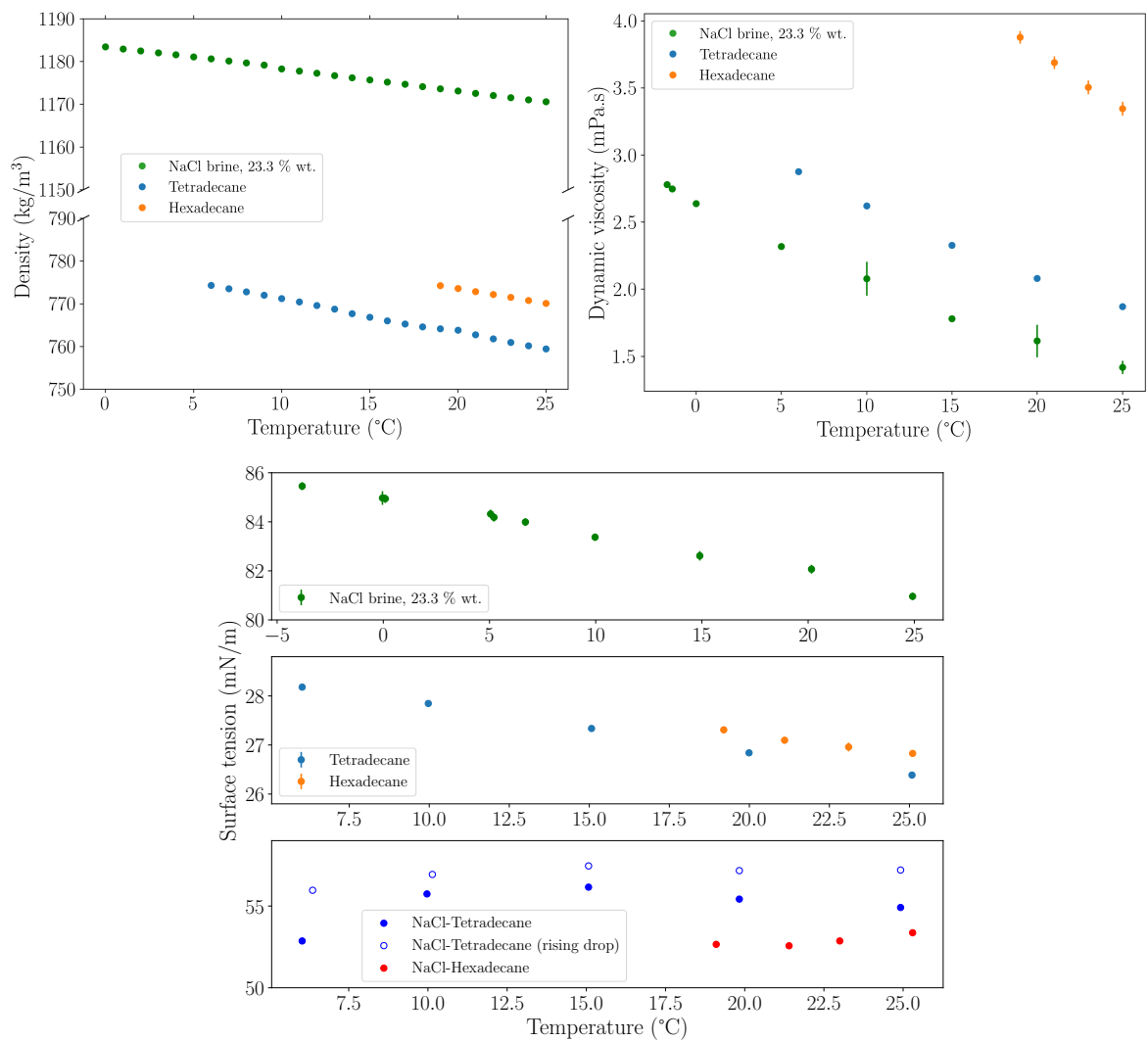

Figure S1: Variations with the temperature of liquid properties, with namely in (a) the density, in (b) the dynamics viscosity and in (c) the surface tension. For the surface tension, the two first plots are liquid-vapor surface tensions and the last one is the liquid-liquid surface tension.

## 1.3 Values used in the present study

For the computed values, the alkane properties in liquid phase are taken at $T_\mathrm{d} = 20$ °C and in solid phase at the melting temperature $T_\mathrm{m}$. As the bath temperature ranges from $T_\mathrm{d}$ to $-21$ °C, the NaCl brine properties are taken at a mean temperature of around 0 °C with an uncertainty to account for the temperature variation. The literature and calculated values are presented in Table S1.

|  | $T_\mathrm{melt}$ (°C) | Density (kg/m³) | | Surface tension (mN/m) | Dynamic viscosity (mPa.s) |
|---|---|---|---|---|---|
|  |  | Liquid $\rho_\mathrm{d}$ | Solid $\rho_\mathrm{s}$ | $\sigma_\mathrm{d}$ | $\mu_\mathrm{d}$ |
| Tetradecane | 5.6 | 763 | 880 | 26.9 | 2.1 |
| Hexadecane | 18.1 | 743 | 886 | 27.2 | 3.8 |
|  |  | $\rho_\mathrm{b}$ | | $\sigma_\mathrm{b}$ | $\mu_\mathrm{b}$ |
| NaCl 23.3 %wt. | $-21$ | $1184 \pm 10$ | | $85 \pm 3$ | $2.6 \pm 1.0$ |

|  | Latent heat (kJ/kg) | Specific heat (kJ/kg − K) | | Conductivity (W/m − K) | | Diffusivity $\times 10^{-7}$ (m²/s) | | Effusivity (W$\sqrt{s}$/m² − K) | |
|---|---|---|---|---|---|---|---|---|---|
|  | $\mathcal{L}$ | $C_\mathrm{p,d}$ | $C_\mathrm{p,s}$ | $\lambda_\mathrm{d}$ | $\lambda_\mathrm{s}$ | $D_\mathrm{d}$ | $D_\mathrm{s}$ | $e_\mathrm{d}$ | $e_\mathrm{s}$ |
| Tetradecane | 227 | 2.20 | 1.80 | 0.14 | 0.14 | 0.83 | 0.88 | 485 | 470 |
| Hexadecane | 236 | 2.20 | 1.81 | 0.14 | 0.22 | 0.86 | 1.37 | 478 | 594 |
|  |  | $C_\mathrm{p,b}$ | | $\lambda_\mathrm{b}$ | | $D_\mathrm{b}$ | | $e_\mathrm{b}$ | |
| NaCl 23.3 %wt. |  | 3.3 | | 0.55 | | 1.41 | | 1466 | |

Table S1: Physical properties of alkanes and NaCl 23.3 %wt. brine used in the study. Values in blue are taken from our measurements, those in orange are computed through equations S1 and S2, and values in black are taken from the literature [1, 2, 3, 4, 5, 6]. The alkane properties are taken at $T_\mathrm{d} = 20$ °C, except the solid properties taken at their respective melting temperature $T_\mathrm{m}$. The NaCl brine properties are taken at a mean temperature of around 0 °C.

## 1.4 Variation of bath viscosity with temperature

The variation of the bath viscosity is plotted in figure S1 for a temperature range of $[-1, 25]$ °C. By extrapolating our data, we expect that the brine viscosity is about 4 mPa.s at $-20$ °C.

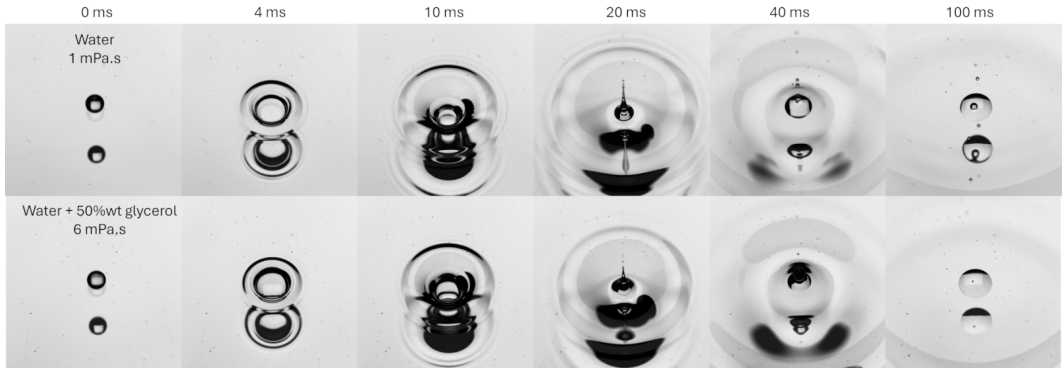

Figure S2: Times series showing the impact of a drop of hexadecane on a bath of pure water (top) and a mixture of water and glycerol at a 50 % wt. ratio (bottom). Both liquids are at room temperature, and have a viscosity of 1 and 6 mPa.s respectively.

Figure S2 represents image sequences of two experiments at room temperature: the top row is an hexadecane drop impacting a pure water bath of viscosity 1 mPa.s and the bottom row is an hexadecane drop impacting a solution of pure water and glycerol at 50 % wt, which has a viscosity of 6 mPa.s. We do not observe a significant change of the impact dynamic, which justifies our choice to neglect the temperature variation of dynamic viscosity of the bath in our study.



\end{document}